\documentclass[copyright,creativecommons]{eptcs}
\usepackage{makeidx}  
\usepackage{breakurl}             
\usepackage[latin1]{inputenc}
\usepackage{amssymb}
\usepackage{upgreek}
\usepackage{graphicx}
\usepackage{pdfsync}
\usepackage{xspace}
\usepackage{theorem}
\newtheorem{DEF}{Definition}[section]
\newtheorem{THM}[DEF]{Theorem}
\newtheorem{LEM}[DEF]{Lemma}
\newtheorem{PROP}[DEF]{Proposition}

\theorembodyfont{\normalfont}

\newcommand{\BREAK}{\hspace*{0mm}\\}

\newcommand{\pp}[1][p]{\ensuremath{\texttt{#1} }}
\newcommand{\hup}{\ensuremath{\hspace{-1mm}\upharpoonright\hspace{-1mm}}}

\newcommand{\llangle}{\ensuremath{\langle\!\langle}}
\newcommand{\rrangle}{\ensuremath{\rangle\!\rangle}}
\newcommand{\llpar}{\ensuremath{(\!(}}
\newcommand{\rrpar}{\ensuremath{)\!)}}
\newcommand{\BOOL}{\keyword{bool}}
\newcommand{\INT}{\keyword{int}}
\newcommand{\DOM}{\keyword{dom}}

\newcommand{\END}{\keyword{end}}
\newcommand{\text}{\textrm}
\newcommand{\AMP}{\ensuremath{\hspace*{0.7mm}\&\hspace*{0.7mm}}}

\newcommand{\lbl}[1]{\textsf{#1}}

\newif\ifrh
\rhtrue
\newif\ifny
\nytrue

\newcommand{\keyword}[1]{\textsf{\upshape #1}\xspace}
\newcommand{\defk}{\keyword{def}}
\newcommand{\ink}{\keyword{in}}
\newcommand{\andk}{\keyword{and}}

\newcommand{\truek}{\keyword{true}}
\newcommand{\falsek}{\keyword{false}}

\newcommand{\ifk}{\keyword{if}}
\newcommand{\thenk}{\keyword{then}}
\newcommand{\elsek}{\keyword{else}}
\newcommand{\notk}{\keyword{not}}

\newcommand{\participant}[1]{\mathtt{#1}}

\newcommand{\IF}[2]{\ifk\ #1\ \thenk\ #2\ \elsek\ }

\newcommand{\branch}{\&}

\newcommand{\typeconst}[1]{\keyword{#1}}

\newcommand{\End}{\typeconst{end}}

\newcommand{\rcdt}{\{l_i\colon T_i\}_{i\in I}}
\newcommand{\brancht}[1][\alpha]{\branch\rcdt}

\newcommand{\VEC}{\tilde}

\def\fps@figure{tp}      
\def\fps@table{tp}
\usepackage{infer}
\usepackage{subfigure}
\usepackage{booktabs}
\usepackage{mathdots}
\usepackage{stmaryrd}
\usepackage{listings}
\usepackage[usenames,dvipsnames]{color}
\usepackage{wrapfig}
\usepackage{float}
\usepackage{slashbox}
\usepackage{space}
\usepackage{placeins}

\newcommand{\Section}[1]{\section{#1} \setcounter{figure}{0}}
\floatstyle{ruled}
\restylefloat{table}
\restylefloat{figure}
\definecolor{darkblue}{rgb}{0,0,0.5}
\definecolor{darkgreen}{rgb}{0,0.5,0}
\lstdefinelanguage{apims_math}
                  {morekeywords={if,then,else,true,false,end,def,String,Int,Bool,Unit,!,?},
                   literate={<}{$\langle$}1
                            {>}{$\rangle$}1
                            {=>}{$\to$}1
                            {rec}{$\upmu$}1
                            {<<}{!}1
                            {>>}{?}1
                            {<:}{$\triangleleft$}1
                            {:>}{$\triangleright$}1
                            {^}{\texttt{\underline{\hspace{5mm}}\hspace{-5mm}} }{-1}
                            {\#}{{}}{-1}
                            {$}{$\!$}1
                            {£}{$\!$}1
                            {|}{$\hup$}1
                            {||}{$\mid$}1
                            {rand}{\tt{rand}}3
                            {sync}{\tt{sync}}3
                            {if}{\tt{if}}2
                            {then}{\tt{then}}3
                            {else}{\tt{else}}3
                            {end}{\sf{end}}2
                            {true}{\sf{true}}3
                            {false}{\sf{false}}3
                            {eData}{$\mathit{eData}$}5
                            {eSchedule}{$\mathit{eSchedule}$}8
                            {eResult}{$\mathit{eResult}$}6
                            {CaseD}{{\sf{CaseD}}{}}5
                            {CaseDD}{\sf{CaseDD}{}}6
                            {CaseDN}{\sf{CaseDN}{}}6
                            {CaseN}{\sf{CaseN}}5
                            {CaseND}{\sf{CaseND}}6
                            {CaseNN}{\sf{CaseNN}}6
                            {^Pdata:}{\sf{\underline{Pdata}:}}5
                            {^Dschedule:}{\sf{\underline{Dschedule}:}}8
                            {^Nschedule:}{\sf{\underline{Nschedule}:}}8
                            {^Dresult:}{\sf{\underline{Dresult}:}}6
                            {\#CaseD:}{\sf{CaseD:}}5
                            {^CaseD:}{\sf{\underline{CaseD}:}}5
                            {\#CaseDD:}{\sf{CaseDD:}}6
                            {^CaseDD:}{\sf{\underline{CaseDD}:}}6
                            {\#CaseDN:}{\sf{CaseDN:}}6
                            {^CaseDN:}{\sf{\underline{CaseDN}:}}6
                            {\#CaseN:}{\sf{CaseN:}}5
                            {^CaseN:}{\sf{\underline{CaseN}:}}5
                            {\#CaseND:}{\sf{CaseND:}}6
                            {^CaseND:}{\sf{\underline{CaseND}:}}6
                            {\#CaseNN:}{\sf{CaseNN:}}6
                            {^CaseNN:}{\sf{\underline{CaseNN}:}}6
                            {P_P}{${\color{red}P_P}$}2
                            {P_D}{${\color{red}P_D}$}2
                            {P_N}{${\color{red}P_N}$}2
                            {P_C}{${\color{red}P_C}$}2
                            {P_P'}{${\color{red}P^\prime_P}$}2
                            {P_D'}{${\color{red}P^\prime_D}$}2
                            {P_N'}{${\color{red}P^\prime_N}$}2
                            {P_C'}{${\color{red}P^\prime_C}$}2
                            {/a}{$\mathtt{\color{darkblue}\overline{a}}$}1
                            ,
                   sensitive=true,
                   moredelim=[is][\rmfamily]{\%}{\%}
                   morecomment=[l]{//},
                   basicstyle=,
                   keywordstyle=\color{darkgreen},
                   identifierstyle=\color{darkblue},
                   commentstyle=\color{Gray},
                   stringstyle=\ttfamily,
                   showstringspaces=true,
                   keywordsprefix={\^,\#}
                  }
\lstdefinelanguage{apims}
                 {morekeywords={true,false,end,Gend,Lend,def,in,link,sync,guisync,String,Int,Bool,Unit,<<,>>},
                   literate=,
                   sensitive=true,
                   morecomment=[l]{//},
                   basicstyle=\footnotesize\ttfamily,
                   keywordstyle=\color{darkgreen}\bfseries,
                   identifierstyle=\color{darkblue}\bfseries,
                   commentstyle=\color{Gray},
                   stringstyle=\ttfamily,
                   showstringspaces=true,
                   keywordsprefix={\^,\#}
                  }
\providecommand{\event}{EXPRESS 2010} 

\title{Multiparty Symmetric Sum Types}
\author{Lasse Nielsen
\institute{DIKU, University of Copenhagen}
\and Nobuko Yoshida
\institute{Imperial College London}
\and Kohei Honda
\institute{Queen Mary, University of London}
}
\def\titlerunning{Multiparty Symmetric Sum Types}
\def\authorrunning{L. Nielsen, N. Yoshida \& K. Honda}
\begin{document}
\maketitle

\begin{abstract}
This paper introduces a new theory of multiparty session types based on
symmetric sum types, by which we can type non-deterministic orchestration
choice behaviours.  While the original branching type in session types can
represent a choice made by a single participant and accepted by others
determining how the session proceeds, the symmetric sum type represents a
choice made by agreement among all the participants of a session. Such
behaviour can be found in many practical systems, including collaborative
workflow in healthcare systems for clinical practice guidelines (CPGs).
Processes using the symmetric sums can be embedded into the original branching
types using conductor processes.  We show that this type-driven embedding
preserves typability, satisfies semantic soundness and completeness, and meets
the encodability criteria \cite{PalamidessiC:comepsapc,GORLA08} adapted to the
typed setting.  The theory leads to an efficient implementation of a
prototypical tool for CPGs which automatically translates the original CPG
specifications from a representation called the Process Matrix to symmetric
sum types, type checks programs and executes them.

\end{abstract}
\vspace{-5mm}
\Section{Introduction}
\label{sec:introduction}
\newcommand{\SYNC}[1][{}]{\ensuremath{\texttt{sync}_{#1} }}
\newcommand{\RAND}{\ensuremath{\texttt{rand} }}
\newcommand{\picalc}{\ensuremath{\uppi\textrm{-calculus}} }
Clinical Practice Guidelines (CPGs) \cite{cpg09} are detailed
descriptions of medical treatment procedures, practised globally with
local variations, in order to treat specific medical disorders.
CPGs are an example of social interactions,
which include workflow models and various cooperation models: its
richness stems from the diverse collaborative patterns human
organisations can exhibit. One such pattern, which plays a prominent
role in CPGs, is {\em symmetric synchronisation} where all the
participants are equal in the decision-making, i.e. the participants
collectively decide on one of the possible choices.

\newcommand{\Participant}[1]{\textit{#1}}
\newcommand{\parP}{\Participant P}
\newcommand{\parD}{\Participant D}
\newcommand{\parN}{\Participant N}
\begin{figure}[b] 
\caption{Cases in the healthcare cooperation example}
\label{fig:example_cases}
\begin{center}
\begin{tabular}{p{7cm}p{2.3cm}p{5cm}}
\vspace{-3.5mm}
\begin{tabular}[t]{|l|c|c|c|} 
\hline
        & Data & Schedule & Inspect \\
\specialrule{0.5mm}{0mm}{0mm}
Case DD & \parD& \parD    & \parD \\
\hline
Case ND & \parN& \parD    & \parD \\
\hline
Case DN & \parD& \parN    & \parD \\
\hline
Case NN & \parN& \parN    & \parD \\
\hline
\end{tabular} 
&
\vspace*{3mm}
\parD: Doctor \newline
\parN: Nurse \newline
\parP: Patient
&
\vspace*{3mm}
Data: Obtain patient data \newline
Schedule: Schedule inspection \newline
Inspect: Perform inspection
\end{tabular}
\end{center}
\vspace{-3mm}
\label{fig:matrix}
\end{figure} 
Motivated from practice, this paper aims to distill the essence of this
symmetric synchronisation as an interaction primitive, position it as part of
the type theory for the asynchronous $\picalc$ with multiparty sessions, and
explore its properties to model workflow frameworks, enjoying the richness of
multiparty session types to express how data is exchanged.
Our starting point is a widely known semi-formal modelling framework for CPGs
and other workflows called Process Matrix \cite{pm08example}, which provides a
concise and general description of symmetric synchronisation patterns as found
in CPGs.

The new synchronisation primitive is generally useful, also for other calculi and applications.
We add the symmetric synchronisation primitive to the asynchronous $\picalc$ and study it in a typed
setting because it allows us to model CPGs as types, and enables correctness
and erasure properties.


We explain the key ideas of Process Matrix and CPGs using an example from a CPG
with three participants: A doctor (\parD), a nurse (\parN) and a patient
(\parP).  The doctor and the nurse need to register and inspect the patient,
thus they must obtain the patient data (Data), schedule an appointment
(Schedule) and inspect the patient (Inspect).
The actions can be divided between the doctor and the nurse in four different
ways, since they both can collect the data and schedule the appointment but
only the doctor may inspect the patient. The four cases are illustrated in the
table in Fig.~\ref{fig:example_cases}. For example in Case ND, the nurse
obtains the patient data and the doctor schedules and performs the
inspection.  In this way, the doctor and the nurse need to perform a
different combination of actions depending on which case is chosen,
thus they need to commit to the same choice, in order for the
cooperation to work.  This choice cannot be implemented directly using the asymmetric
choice (as found in branching/selection primitives in the foregoing
session types \cite{THK,honda.vasconcelos.kubo:language-primitives}),
since the decision would be done by a single participant and not by common
agreement.

Our aim is to obtain a general modelling framework which can uniformly
capture both symmetric synchronisations and existing session-based
communication patterns. Such a framework will give a basis for the
implementation of a tool for CPGs where one can describe, validate and
execute specifications backed up by static validation coming from the
theory.
For this purpose we incorporate the synchronisation primitive in the
type theory for multiparty sessions from \cite{BC07,CHY08}, so different
groups of principals freely can mix standard asymmetric communications and
symmetric synchronisations. The resulting sessions are abstracted as types,
enabling type-based validation which ensures type and communication safety.

We offer the first prototype implementation of the $\picalc$ with multiparty
sessions, with a typechecker using multiparty session types with full
projections. Our implementation includes the symmetric synchronisation
primitive and verification using symmetric sum types. This allows us to implement,
verify and execute the examples used to explain and motivate the extension.

The use of types is not only essential for modelling CPGs and validating processes,
but also enables an organised analysis of the synchronisation
primitive. Using a type-directed translation, we show that the primitive can
be embedded into the asymmetric branching in the original multiparty sessions
\cite{BC07,CHY08}.  The translation generates auxiliary
processes from the types, and combines them with an encoding
of the sum into asymmetric branch types, respecting global interaction
patterns and preserving semantics, by exploiting the type structure. The
auxiliary process generated from a type conducts the synchronisations of a session by receiving
accepted cases from participants and sending the chosen case
back.  To prove its correctness, we use a new technique based on 
derivations of the multiparty session typing. 
The resulting translation introduces
exponentially more branching cases (e.g.~64 for the running example),
demonstrating the practical usefulness of the symmetric sum for
compact description as well as offering a formally founded distributed
implementation strategy of the primitive.

Next we present the calculus for
multiparty symmetric synchronisation (Section~\ref{sec:processes}) and
study its type theory (Section~\ref{sec:typesystem}). We then define a
type-directed encoding (Section~\ref{sec:erasure}) of the symmetric
sum into the asynchronous multiparty session; and investigate its
encodability criteria by adapting the framework from
\cite{PalamidessiC:comepsapc,GORLA08} to the typed setting.  Finally
we present an application of the theory to the formal CPG
verification (Section~\ref{sec:verification}), 
with a prototype implementation available from \cite{apims:projectpage}.
The technical contributions include
\emph{subject reduction} (Theorem~\ref{thm:subject_reduction})
and \emph{type/semantic correctness} of the
encoding (Theorems~\ref{thm:erasure:typable}, \ref{thm:ext:5221} and
\ref{thm:erasure:completeness}). 
The implementation demonstrates the correctness,
feasible implementability and significance of the new primitive.
In particular, an automatic mapping
from Process Matrix to global types 
(Section~\ref{sec:verification}) shows the expressiveness 
of multiparty session types.  
Appendix in the full version \cite{sumtypes:appendix} includes the omitted definitions, examples and proofs,
though the paper can be read independently.

\newpage
\Section{Processes with Synchronisation}
\label{sec:processes}
\begin{figure}[t] 
\small
\begin{tabular}{p{6cm}p{2.3cm}p{0.4cm}p{2.8cm}p{2.75cm}}
$P              ::= {\SYNC[\tilde s,n]\{l: P_l\}_{l \in L}}$ \newline
${}\hspace{6mm}\mid {\RAND\{P_i\}_{i \in I}}$ \newline
${}\hspace{6mm}\mid \overline{a}[\participant{2..n}](\tilde s).P $ \newline
${}\hspace{6mm}\mid a[\participant{p}](\tilde s).P $ \newline
${}\hspace{6mm}\mid s!\langle \tilde e \rangle;P $ \newline
${}\hspace{6mm}\mid s?(\tilde x);P $ \newline
${}\hspace{6mm}\mid s!\llangle\tilde s\rrangle;P $ \newline
${}\hspace{6mm}\mid s?\llpar\tilde s\rrpar;P $ \vspace*{1mm}\newline
\newline
$\nobreak{D        ::= \{X_i(\tilde x_i \tilde s_i) = P_i\}_{i \in I} }$ \newline
$e                 ::= v \mid x \mid e\ \andk\ e' \mid \notk\ e \mid \RAND\{v_i\}_{i \in I} \mid ... $
&
$ $ \hfill synchronisation \newline
$ $ \hfill random choice \newline
$ $ \hfill session request \newline
$ $ \hfill session accept \newline
$ $ \hfill value sending \newline
$ $ \hfill value reception \newline
$ $ \hfill delegation \newline
$ $ \hfill reception \vspace*{1mm}\newline
\newline
$ $ \hfill declarations \newline
$ $ \hfill expressions \newline
&
\quad \quad \quad 
&
$\mid s\triangleleft l;P$ \newline
$\mid s\triangleright \{l: P_l\}_{l \in L}$ \newline
$\mid {\IF e P Q}$ \newline
$\mid P|Q $ \newline
$\mid 0 $ \newline
$\mid (\nu n)P $ \newline
$\mid \defk\ D\ \ink\ P $ \newline
$\mid X \langle \tilde e \tilde s\rangle $ \newline
$\mid s: \tilde h $ \vspace*{1mm}
\newline
$v                 ::= a \mid \truek \mid \falsek $ \newline
$h                 ::= l \mid \tilde v \mid \tilde s$ \newline
&
$ $ \hfill label selection \newline
$ $ \hfill label branching \newline
$ $ \hfill conditional \newline
$ $ \hfill parallel \newline
$ $ \hfill inaction \newline
$ $ \hfill restriction \newline
$ $ \hfill recursion \newline
$ $ \hfill process call \newline
$ $ \hfill message queue \vspace*{1mm}
\newline
$ $ \hfill values \newline
$ $ \hfill messages \newline
\end{tabular}
\vspace{1mm}
\caption{The process language}
\label{fig:processlang}
\vspace*{-5mm}
\end{figure} 
\begin{figure}[b] 
\small
\begin{center}
\Infer[Link]{
\overline{a}[2..n](\tilde s).P_1 | a[2](\tilde s).P_2 | \ldots | a[n](\tilde s).P_n
   \to (\nu \tilde s)(P_1 | P_2 | \ldots | P_n | s_1:\emptyset | \ldots | s_m:\emptyset)}
      {}
\hspace{5mm}
\Infer[Send]{
s!\langle \tilde e \rangle;P | s:\tilde h
   \to P | s:\tilde h \cdot \tilde v}
      {\tilde e \downarrow \tilde v}
\\[1mm]
\Infer[Recv]{
s?(\tilde x);P | s: \tilde v \cdot \tilde h
   \to P[\tilde v/\tilde x] | s:\tilde h}
      {} 
\hspace{5mm}
\Infer[Label]{
s \triangleleft l;P | s:\tilde h
   \to P | s:\tilde h \cdot l}
      {} 
\hspace{5mm}
\Infer[Branch]{
s\triangleright \{l_i: P_i\}_{i \in I} | s: l_j \cdot \tilde h
   \to P_j | s:\tilde h}
      {j \in I} 
\\[1mm]
\Infer[Deleg]{
s!\llangle \tilde t \rrangle;P | s:\tilde h
   \to P | s:\tilde h \cdot \tilde t}
      {}
\hspace{5mm}
\Infer[SRec]{
s?\llpar\tilde t\rrpar;P | s:\tilde t \cdot \tilde h
   \to P | s:\tilde h}
      {} 
\hspace{5mm}
\Infer[IfT]{
\IF e P Q  \to P}
      {e \downarrow \truek} 
\\[1mm]
\Infer[IfF]{
\IF e P Q  \to Q}
      {e \downarrow \falsek} 
\hspace{5mm}
\Infer[Def]{
\defk\ D\ \ink\ X\langle \tilde e \tilde s \rangle | Q \to \defk\ D\ \ink\ P[\tilde v/\tilde x] | Q}
      {\tilde e \downarrow \tilde v
 \quad X\langle \tilde x \tilde s \rangle = P \in D} 
\hspace{5mm}
\Infer[Scop]{
(\nu n)P \to (\nu n) P'}
      {P \to P'} 
\\[1mm]
\Infer[Par]{
P | Q \to P' | Q}
      {P \to P'} 
\hspace{5mm}
\Infer[Defin]{
\defk\ D\ \ink\ P \to \defk\ D\ \ink\ P'}
      {P \to P'} 
\hspace{5mm}
\Infer[Str]{
P' \to Q'}
      {P \equiv P'
\quad  P \to Q
\quad  Q \equiv Q'} 
\\[1mm]
\Infer[Rand]{\RAND\{P_i\}_{i \in I} \to P_j}
      {j \in I}
\hspace{5mm}
\Infer[Sync]{\SYNC[\tilde s,n]\{l: P_{1l}\}_{l \in L_1} \mid ... \mid \SYNC[\tilde s,n]\{l: P_{nl}\}_{l \in L_n} \to P_{1h} \mid ... \mid P_{nh} }
      {h \in \bigcap_{i=1}^n L_i}
\end{center}
\caption{The reduction rules}
\label{fig:semantics}
\vspace*{-3mm}
\end{figure} 
This section introduces the syntax (Fig.~\ref{fig:processlang}) of the
asynchronous multiparty session $\picalc$ \cite{CHY08} with the new
$\mathtt{sync}$ primitive, and the judgement $P \to P'$
(Fig.~\ref{fig:semantics}, where $e \downarrow v$ denotes the evaluation of the
expression $e$ to the value $v$) describing the small-step semantics for
processes.   
The syntax defines the values: $\{v,w,\ldots\}$, expressions: $\{e,e',\ldots\}$
and processes: $\{P,Q,\ldots\}$ from the sets of channel names:
$\{a,b,\ldots\}$, value variables: $\{x,y,\ldots\}$, session channels:
$\{s,t,\ldots\}$, labels: $\{l, m, \ldots\}$ and process variables:
$\{X,Y,\ldots\}$.

Session request, $\overline{a}[\participant{2..n}](\tilde s).P$ initiates a session
with channels $\tilde s$ (where $\tilde s$ denotes a vector $s_1\ldots s_n$)
over the public channel $a$ with the other $n-1$
participants of shape $a[\pp](\tilde s).Q_{\pp}$ for
$\pp$ from $2$ to $n$ ({\sc\small [Link]} in Fig.~\ref{fig:semantics}). 
Asynchronous communication in an established session is performed by sending
and receiving values ({\sc\small [Send,Recv]}), transferring a session 
using session delegation and reception ({\sc\small [Deleg,SRec]}), 
and label selection and branching ({\sc\small [Label,Branch]}),
where the branching process offers a number of labels and the selecting process
chooses one of them.

The new $\SYNC[\tilde s,n]\{l: P_l\}_{l \in L}$ constructor is interpreted as
the process participating in a plenum decision between all the $n$ processes in
the session $\tilde s$ reaching a common decision $h$ from $L$.
Afterwards the process proceeds as described in $P_h$.  In {\sc\small
[Sync]} in Fig.~\ref{fig:semantics},  $h$ in the premise denotes the common label.    
We also add the $\RAND\{P_i\}_{i \in I}$ constructor which randomly
selects one of its branches ({\sc\small [Rand]}). 
This primitive can be expressed using $\ifk$ and a random expression
(hence it does not add expressiveness from \cite{CHY08}),
but simplifies the erasure mapping in Section~\ref{sec:erasure}.

In {\sc\small [Sync]},   the processes cannot perform the synchronisation if
they do not share some common label, in which case  the processes will be
stuck.  We also need to know how many participants are in the session in order
to know when the synchronisation can step; otherwise the processes will be
stuck.  The typing system introduced in the next section ensures that $\SYNC$
satisfies these two conditions.  

\begin{figure}[t] 
\vspace{-4mm}
\scriptsize
\begin{tabular}{p{8.8cm}p{6.2cm}}
\begin{lstlisting}
P_D = // Doctor
a[2](d,s,r,cp,cn).
if rand{true, false}
then
 cp<:CaseD;cn<:CaseD; d>>(data);
 if rand{true, false}
 then 
  cp<:CaseDD;cn<:CaseDD; s<<<eSchedule>;r<<<eResult>;end
 else
  cp<:CaseDN;cn<:CaseDN; r<<<eResult>;end
else
 cp<:CaseN;cn<:CaseN;
 if rand{true, false}
 then
  cp<:CaseND;cn<:CaseND; s<<<eSchedule>;r<<<eResult>;end
 else
  cp<:CaseNN;cn<:CaseNN; r<<<eResult>;end
\end{lstlisting}
&
\begin{lstlisting}
P_P = // Patient
/a[2..3](d,s,r,cp,cn). pd:>
{CaseD: d<<<eData>; cp:>
  {CaseDD: s>>(schedule);r>>(result);end,
   CaseDN: s>>(schedule);r>>(result);end },
 CaseN: d<<<eData>; pd:>
  {CaseND: s>>(schedule);r>>(result);end,
   CaseNN: s>>(schedule);r>>(result);end }
}
P_N = // Nurse
a[3](d,s,r,cp,cn). cn:>
{CaseD: cn:>
  {CaseDD: end,
   CaseDN: s<<<eSchedule>;end },
 CaseN: d>>(data); cn:>
  {CaseND: end,
   CaseNN: s<<<eSchedule>;end } }
\end{lstlisting}
\end{tabular}
\vspace*{-5mm}
\caption{Healthcare Example without $\mathtt{sync}$}
\label{fig:example_nosync}
\end{figure} 
\begin{figure}[b] 
\vspace*{-4mm}
\scriptsize
\begin{tabular}{p{20cm}}
\begin{lstlisting}
P_P = // Patient
/a[2..3](d,s,r). sync((d,s,r),3)
{#CaseD: d<<<eData>; sync((d,s,r),3)
  {#CaseDD: s>>(schedule);r>>(result);end, #CaseDN: s>>(schedule);r>>(result);end },
 #CaseN: d<<<eData>; sync((d,s,r),3)
  {#CaseND: s>>(schedule);r>>(result);end, #CaseNN: s>>(schedule);r>>(result);end } }
\end{lstlisting}
\vspace*{-3mm}
\begin{lstlisting}
P_D = // Doctor
a[2](d,s,r). sync((d,s,r),3)
{#CaseD: d>>(data); sync((d,s,r),3) {#CaseDD: s<<<eSchedule>;r<<<eResult>;end, #CaseDN: r<<<eResult>;end },
 #CaseN:           sync((d,s,r),3) {#CaseND: s<<<eSchedule>;r<<<eResult>;end, #CaseNN: r<<<eResult>;end } }
\end{lstlisting}
\vspace*{-3mm}
\begin{lstlisting}
P_N = // Nurse
a[3](d,s,r). sync((d,s,r),3)
{#CaseD:           sync((d,s,r),3) {#CaseDD: end, #CaseDN: s<<<eSchedule>;end },
 #CaseN: d>>(data); sync((d,s,r),3) {#CaseND: end, #CaseNN: s<<<eSchedule>;end } }
\end{lstlisting}
\end{tabular}
\vspace*{-5mm}
\caption{Healthcare Example using $\mathtt{sync}$}
\label{fig:example_sync}
\end{figure} 
\paragraph{Healthcare Cooperation (1): Processes}
We motivate the symmetric synchronisation using the example from the
introduction.  We first explain the problem when representing this
interaction without $\SYNC$.  As explained in the introduction, there is no
rigorous way to decide which of the four cases will occur, as well as
who will be the principal decision maker: we could let the doctor
non-deterministically decide between the cases, and then we obtain the processes in
Fig.~\ref{fig:example_nosync}, if we are to use the processes from
\cite{CHY08}: similarly we could let the nurse or even the patient
decide. None of these representations captures the cooperation where
the doctor, the nurse and the patient should reach a common decision,
because {\em it is impossible to know who takes the initiative}.
Another problem is that we need to specify the choices in $P_D$, which
is best captured by non-deterministic expressions like \keyword{rand}.

Fig.~\ref{fig:example_sync} describes the same example using $\SYNC$ where the
intended cooperation is directly modelled.  The case is logically decided by
two choices: first it is decided who receives the patient data, and then it is
decided who schedules the inspection. Since these decisions are not
necessarily made at the same time, the processes select the case using two
sequential synchronisations.

\Section{Symmetric Sum Types}
\label{sec:typesystem}
\begin{figure}[t] 
\small
\vspace{1mm}
\hspace*{5mm}
\begin{tabular}{p{5cm}p{5cm}p{5cm}}
\hspace*{-0.5cm}(Global Types) \newline 
$G               ::= \pp \to \pp': k \langle U \rangle.G' $ \newline
${}\hspace{6mm} \mid \pp \to \pp': k \{l_i: G_i\}_{i \in I} $ \newline
${}\hspace{6mm} \mid \upmu t.G \mid t \mid \END $ \newline
${}\hspace{6mm} \mid \{l: G_l\}_{l \in L;M}\ (M \neq \emptyset)$
&
\hspace*{-0.5cm}(Local Types) \newline
$T               ::= k!\langle U \rangle;T $ \newline
${}\hspace{5.5mm} \mid k?\langle U \rangle;T $ \newline
${}\hspace{5.5mm} \mid k\oplus \{l: T_l\}_{l \in L} $ \newline
${}\hspace{5.5mm} \mid k\AMP\{l: T_l\}_{l \in L} $ \newline
${}\hspace{5.5mm} \mid \upmu t.T \mid t \mid \END $ \newline
${}\hspace{5.5mm} \mid \{l: T_l\}_{l \in L;M}\ (M \neq \emptyset)$
&
\hspace*{-0.5cm}(Message Types) \newline 
$U               ::=  \tilde S \mid T@(\pp,m,n) $ \newline
\hspace*{-0.5cm}(Simple Types) \newline
$S               ::= \BOOL \mid \INT \mid ... \mid \langle G \rangle $ \newline
\hspace*{-0.5cm}(Environments) \newline
$\Gamma ::= \emptyset \mid \Gamma, u: \langle G \rangle \mid \Gamma, X: \tilde S \tilde T$ \newline
$\Delta ::= \emptyset \mid \Delta, \tilde s: T@(\pp,n)$
\end{tabular}
\caption{The Domains used for Global and Local types}
\label{fig:globaltypes}
\label{fig:localtypes}
\end{figure} 
\begin{figure}[t] 
\caption{Selected typing rules}
\label{fig:typingrules}
\small
\begin{center}
\Infer[Rand]
     {\Gamma \vdash \RAND \{P_i\}_{i \in I} \rhd \Delta}
     {\forall i \in I. \Gamma \vdash P_i \rhd \Delta}
\quad
\Infer[Sync]
     {\Gamma \vdash \SYNC[\tilde s,n]\{l: P_l\}_{l \in L''} \rhd \Delta, \tilde s: \{l: T_l\}_{l \in L;L'}@(\pp,n)}
     {\forall l\in L'': \Gamma \vdash P_l \rhd \Delta, \tilde s: T_l@(\pp,n)
\quad L'' \subseteq L \cup L'
\quad L' \subseteq L''
     } \\
\vspace{2mm}
\Infer[Mcast]
     {\Gamma \vdash \overline a [\pp[2..n]](\tilde s).P \rhd \Delta}
     {\Gamma \vdash a: \langle G \rangle
\quad \Gamma \vdash P \rhd \Delta,\tilde s:(G \hup 1)@(\pp[1],n)
\quad |\VEC{s}|=\keyword{max}(\keyword{sid}(G))
\quad n=\keyword{max}(\keyword{pid}(G))
     }\\
\vspace{2mm}
\Infer[Macc]
     {\Gamma \vdash a[\pp[p]](\tilde s).P \rhd \Delta}
     {\Gamma \vdash a: \langle G \rangle
\quad \Gamma \vdash P \rhd \Delta,\tilde s:(G \hup \pp)@(\pp,n)
\quad |\VEC{s}|=\keyword{max}(\keyword{sid}(G))
\quad n=\keyword{max}(\keyword{pid}(G))
     }\\
\vspace{2mm}
\Infer[Send]
     {\Gamma \vdash s_k!\langle\tilde e\rangle;P \rhd \Delta, \tilde s: k!\langle\tilde S\rangle;T@(\pp,n)}
     {\forall j. \Gamma \vdash e_j : S_j
\quad \Gamma \vdash P \rhd \Delta, \tilde s: T@(\pp,n)
     }
\quad
\Infer[Rcv]
     {\Gamma \vdash s_k?(\tilde x);P \rhd \Delta, \tilde s: k?\langle\tilde S\rangle;T@(\pp,n)}
     {\Gamma, \tilde x: \tilde S \vdash P \rhd \Delta, \tilde s: T@(\pp,n)
     }
\\\vspace{2mm}
\Infer[Sel]
     {\Gamma \vdash s_k \triangleleft h;P \rhd \Delta, \tilde s: k \oplus \{l: T_l\}_{l \in L}@(\pp,n)}
     {\Gamma \vdash P \rhd \Delta, \tilde s: T@(\pp,n)
\quad h \in L
     }
\quad
\Infer[Branch]
     {\Gamma \vdash s_k \triangleright \{l: P_l\}_{l \in L} \rhd \Delta, \tilde s: k ? \{l: T_l\}_{l \in L}@(\pp,n)}
     {\forall l \in L:
\quad \Gamma \vdash P_l \rhd \Delta, \tilde s: T_l@(\pp,n)
     }
\\\vspace{2mm}
\MainInfer[Conc]
     {\Gamma \vdash P|Q \rhd \Delta \circ \Delta'}
     {\Gamma \vdash P \rhd \Delta
\quad \Gamma \vdash Q \rhd \Delta'
     }
     {\hspace*{-2mm}($\DOM(\Delta) \cap \DOM(\Delta') = \emptyset$)}
\end{center}
\vspace*{-3mm}
\end{figure} 
We start by defining the global types $G$ in Fig.~\ref{fig:globaltypes},
which specifies global session protocols between the participants. Except for
the symmetric sum type, the syntax is from \cite{CHY08}.
The type $\pp \to \pp': k \langle U \rangle.G'$ expresses that participant $\pp$ sends
a message of type $U$ along channel $k$ to $\pp'$ and then interactions described
in $G'$ take place.
The type $\pp \to \pp': k \{l_i: G_i\}_{i \in I}$ expresses that $\participant{p}$ sends
one of the labels $l_i$ to $\pp'$. If $l_j$ is sent, interactions described in
$G_j$ take place.
Type $\upmu t.G$ is a recursive type, assuming type variables ($t, t', \dots$)
are guarded in the standard way. 
We assume that $G$ in the grammar of sorts is closed, i.e., without free type
variables. Type $\End$ represents the session termination. 

The sum type $\{l: G_l\}_{l \in L;M}$ represents a synchronisation where the
labels are taken from the set $L$ and the non-empty set $M$. The labels in $L$
are optional, but the labels in $M$ are mandatory and must be accepted by all
the participants.  The mandatory labels will be underlined to distinguish them from
the optional labels 
(e.g.~$\{l: G_l\}_{l \in \{l1\};\{l2\}} = \{l1: G_{l1}, \underline{l2}: G_{l2}\}$).

The local types $T$ are defined in Fig.~\ref{fig:localtypes}. They
describe the communication performed by a single process. Therefore
the ``from process to process on channel'' syntax is simply changed to sending
or receiving on a channel. Thus the sending type is $k!\langle U \rangle;T$ and
represents sending a message of type $U$ on channel $k$, followed  by the
communication described by $T$.  The type of receiving is $k?\langle U
\rangle;T$, the type of selecting is $k\oplus \{l: T_l\}_{l \in L}$ and the
type of branching is $k \AMP\{l: T_l\}_{l \in L}$.
The difference from \cite{CHY08} is that the symmetric sum type constructor
$\{l: T_l\}_{l \in L; M}$ is added where $L,M$ satisfies the conditions similar
to those of a global sum type.

The message type $T@(\pp,m,n)$ is used for delegation. It describes an open
session, and includes information about the participant number $\pp$, the
number of session channels $m$, and the number of participants $n$ in the
session together with a local type $T$ describing the remaining communication.

Finally we define the global environment $\Gamma$ containing the global types
for shared channels $u$, and process variables $X$, and the local type
environment $\Delta$ containing the remaining session communication in
Fig.~\ref{fig:globaltypes}, where $\tilde s: T@(p,n)$ means $\tilde s$ is an
open session with $n$ participants, where $T$ describes the remaining
communication for participant $\pp$.

The {\em projection} $G \hup \pp$ of a global type $G$ for a participant $\pp$
generates the local type for the participant in an intuitive way, for example
$(\pp_0 \to \pp_1 : k \langle U \rangle.G') \hup \pp$
becomes $k!\langle U \rangle;(G' \hup \pp)$ if $\pp=\pp_0$ and $\pp \neq \pp_1$.
The differences from the definition in
\cite{CHY08} is that we have added a case for the symmetric sum type,
$(\{l: G_l\}_{l \in L;M}) \hup \pp = \{l: (G_l \hup \pp) \}_{l \in L;M}$.

A global type $G$ is coherent \cite{CHY08} if and only if the projection $G
\hup \pp$ is defined for all participants, and $G$ does not allow racing
conditions (linearity).  We only consider coherent global types.

\paragraph{Judgement}
The typing judgement extends the one from \cite{CHY08} with symmetric sum types.
The judgement $\Gamma \vdash P \rhd \Delta$ states that the process $P$ in
the environment $\Gamma$ performs exactly the session communication described
in $\Delta$.
\newpage
The main rules are included in Fig.~\ref{fig:typingrules}. The local types now
carry information about the number of participants $n$ and channels $m$. The
number of participants and channels is determined at the session
initialisation in the rules {\sc[Mcast]} and {\sc[Macc]}, where $\keyword{sid}(G)$ denotes
channels that appear in $G$ and $\keyword{pid}(G)$ denotes the participants
that appear in $G$.
The rule {\sc [Sync]} checks that the synchronisation uses the correct number
of participants, the accepted branches includes the mandatory ones and does not
exceed the optional ones, and checks that each accepted branch is typed with
the correct communication.  The typing rule {\sc [Rand]} checks that each
choice in a $\RAND$ process has the same session environment.

Since the process is reduced by each rule-application, the typability question
$\Gamma \vdash P \rhd \Delta$ is decidable.

\paragraph{Healthcare Cooperation (2): Types}
\begin{figure}[b] 
\vspace{-3mm}
\scriptsize
\begin{tabular}{p{9cm}p{6cm}}
\begin{lstlisting}
G = // Global type
{#CaseD:
  1=>2:1<Sdata>;
  {#CaseDD: 2=>1:2<Sschedule>;2=>1:3<Sresult>;end,
   ^CaseDN: 3=>1:2<Sschedule>;2=>1:3<Sresult>;end
  },
 ^CaseN:
  1=>3:1<Sdata>;
  {^CaseND: 2=>1:2<Sschedule>;2=>1:3<Sresult>;end,
   #CaseNN: 3=>1:2<Sschedule>;2=>1:3<Sresult>;end
  }
}
\end{lstlisting}
&
\begin{lstlisting}
G|1 = // Local type for Patient
{#CaseD:
  1<<<Sdata>;
  {#CaseDD: 2>><Sschedule>;3>><Sresult>;end,
   ^CaseDN: 2>><Sschedule>;3>><Sresult>;end
  },
 ^CaseN:
  1<<<Sdata>;
  {^CaseND: 2>><Sschedule>;3>><Sresult>;end,
   #CaseNN: 2>><Sschedule>;3>><Sresult>;end
  }
}
\end{lstlisting}
\end{tabular}
\vspace{-4mm}
\caption{Global Type $G$ and Patient Projection for Healthcare Example}
\label{fig:example_synctype}
\end{figure} 
We explain how the types can describe and verify the healthcare scenario in the
Introduction. Recall the processes from Fig.~\ref{fig:example_sync}.
To type $P_P \mid P_D \mid P_N$, we need a matching type-environment first. The
processes use the public channel $a$ to create a session, so the environment
must be of the form $\Gamma = a: \langle G \rangle$ for some global type $G$.

We will start by finding the type describing the interactions in \lbl{CaseND}.
First the participants select the choice \lbl{CaseN} and the patient sends the
data to the nurse.  Then the participants select the choice \lbl{CaseND},
the doctor sends the schedule to the patient, and finally the doctor sends the
result to the patient. 

When the patient
has id $\pp[1]$, the doctor has id $\pp[2]$ and the nurse has id $\pp[3]$ the
described communication for \lbl{CaseND} is described by the type
\begin{lstlisting}
{^CaseN: 1=>3:1<Sdata>. {^CaseND: 2=>1: 2<Sschedule>. 2=>1: 3<Sresult>. end} }
\end{lstlisting}
Performing the same reasoning for \lbl{CaseDD}, \lbl{CaseDN} and \lbl{CaseNN}
and adding their branches to the symmetric sums results in the global type $G$
in Fig.~\ref{fig:example_synctype}.  We select \lbl{CaseND}, \lbl{CaseDN} and
\lbl{CaseN} as the mandatory labels. Since all participants must accept the
mandatory choices, this means that it is always possible for the participants
to agree on a choice in each of the synchronisations.
We can then find the local type for the patient process as the patient's projection
of $G$, given in Fig.~\ref{fig:example_synctype}.
Using this type and the projections we can now typecheck the processes.

\newcommand{\DD}{\ensuremath{\mathcal{D} }}
\begin{PROP} 
\label{prop:example_sync:typed}
$a:\langle G \rangle \vdash P_D \mid P_N \mid P_P \rhd \emptyset$.
\end{PROP} 
%
We end this section by proving \emph{subject reduction}, from which 
we can derive 
soundness, communication safety and progress \cite[\S~5]{CHY08} as
corollaries.
Below $\Delta \to^{0/1} \Delta'$ denotes zero or one step using the type
reduction \cite{CHY08}, which represents the communication between 
dual local types. 
For instance, a reduction between input and output types is defined as:
$k!\langle U\rangle;T_1@(\pp,n), k?\langle U\rangle;T_2@(\pp[q],n) \to T_1@(\pp,n),T_2@(\pp[q],n)$.
We extend it to the symmetric sum as: 
$\{ \{ l: T_{\pp}, \ldots \}@(\pp,n) \}_{\pp \in \{1..n\}}
\to \{T_{\pp}@(\pp,n)\}_{\pp \in \{1..n\}}. $

The formulation uses the extension of the typing to runtime processes
($\Gamma \vdash P \rhd_{\tilde t} \Delta$), which corresponds to the presented
typing on processes without open sessions, but also accept processes with open
sessions. This is obtained by joining compatible session environments ($\Delta,
\Delta'$) using the $\Delta \circ \Delta'$ operation to a single environment
expressing the communication in both $\Delta$ and $\Delta'$.
%
Then we have: 
\begin{THM}[Subject Reduction] 
\label{thm:subject_reduction}
\BREAK
If $\Gamma \vdash P \triangleright_{\tilde s} \Delta$, $\Delta$ 
coherent and $P \to P'$
then $\Gamma \vdash P'\triangleright_{\tilde s} \Delta'$ where $\Delta \to^{0/1} \Delta'$.
\end{THM} 
{\sc Proof}: By induction on the derivation of $P \to P'$.

\Section{From Symmetric Sum to Conducted Branching}
\label{sec:erasure}
\newcommand{\eraseGk}[1]{\ensuremath{\left\llbracket #1 \right\rrbracket}}
\newcommand{\eraseUk}[1]{\ensuremath{\left\llbracket #1 \right\rrbracket}}
\newcommand{\eraseTk}[1]{\ensuremath{\left\llbracket #1 \right\rrbracket}}
\newcommand{\erasePk}[1]{\ensuremath{\mathcal{E}\left\llbracket #1 \right\rrbracket}}
\newcommand{\eraseCk}[1]{\ensuremath{\mathcal{C}\left\llbracket #1 \right\rrbracket}}
\newcommand{\eraseGammak}[1]{\ensuremath{\left\llbracket #1 \right\rrbracket}}
\newcommand{\eraseDeltak}[1]{\ensuremath{\left\llbracket #1 \right\rrbracket}}
\begin{figure}[t] 
\begin{center}
\subfigure[Choice without $\SYNC$]{\includegraphics[width=3.4cm]{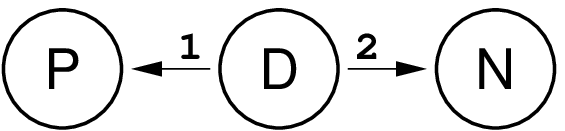}\label{fig:choice:nosync}\hspace*{8mm}}
\subfigure[Choice using $\SYNC$]{\hspace*{4mm}\includegraphics[width=3.4cm]{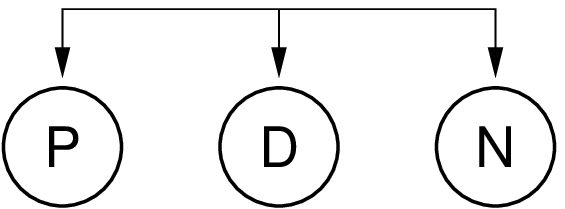}\label{fig:choice:sync}\hspace*{4mm}}
\subfigure[Choice after erasure]{\hspace*{8mm}\includegraphics[width=3.4cm]{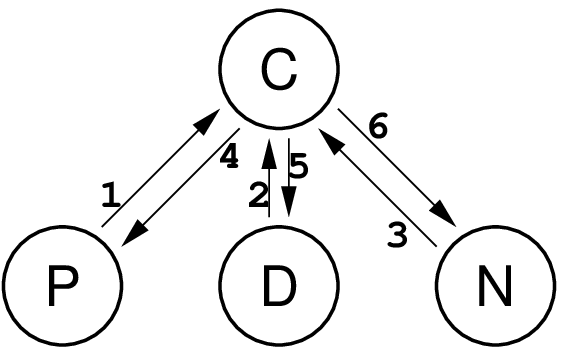}\label{fig:choice:erasure}}
\end{center}
\vspace{-5mm}
\caption{Synchronisation message flows}
\label{fig:choice}
\end{figure} 
This section studies an erasure of symmetric synchronisation, which 
translates away symmetric sums using existing session primitives, which we
hereafter simply call {\em the erasure}. The erasure removes all occurrences of
the $\SYNC$ constructor while preserving static and dynamic semantics, i.e.
typability and reduction. It uses a conductor process for each
session. The messages and protocol used to implement the
synchronisation are illustrated in Fig.~\ref{fig:choice} where the numbers
indicate the sequence of the messages.
Fig.~\ref{fig:choice}(a) shows the communication between the processes
without using $\SYNC$ in Fig.~\ref{fig:example_nosync}.
Fig.~\ref{fig:choice}(b) shows the communication between the processes
using $\SYNC$ in Fig.~\ref{fig:example_sync}, where no messages are sent, because
the synchronisation ensures the same branch is chosen.
Fig.~\ref{fig:choice}(c) shows the conduction messages in the
processes where the synchronisation has been erased in
Fig.~\ref{fig:example_erasure}. First the patient, the doctor and the
nurse send the cases they can accept to the conductor, who chooses a common
case and sends the selected case to the patient, the doctor and the nurse.

\begin{figure}[t] 
\scriptsize
\vspace{1mm}
$\erasePk{
\MainInfer[Mcast]
     {\Gamma \vdash \overline a[\pp[2..n]](\tilde s).P \rhd \Delta}
     {\Gamma \vdash a: \langle G \rangle
\quad \DD_1::\Gamma \vdash P \rhd \Delta,\tilde s:(G \hup \pp[1])@(\pp[1],n)
\quad \begin{array}[b]{c}|\tilde s|=\keyword{max}(\keyword{sid}(G)) 
\\ n=\keyword{max}(\keyword{pid}(G)) \end{array}
     }
  }  =
\begin{array}{l}
  \eraseCk{G}_{\tilde s,n,a} \mid \\
  \overline a[\pp[2..n,n+1]](\tilde s, \texttt{in}_{\tilde s 1},
                                       \texttt{out}_{\tilde s 1}, \ldots,
                                       \texttt{in}_{\tilde s n},
                                       \texttt{out}_{\tilde s n}).\\
  \erasePk{\DD_1}
\end{array}$
\\[2mm]
$\erasePk{
\MainInfer[Macc]
     {\Gamma \vdash a[\pp[p]](\tilde s).P \rhd \Delta}
     {\Gamma \vdash a: \langle G \rangle
\quad \DD_1::\Gamma \vdash P \rhd \Delta,\tilde s:(G \hup \pp)@(\pp,n)
\quad \begin{array}[b]{c}|\tilde s|=\keyword{max}(\keyword{sid}(G)) 
\\ n=\keyword{max}(\keyword{pid}(G)) \end{array}
     }
  }  = a[\pp](\tilde s, \texttt{in}_{\tilde s1}, \texttt{out}_{\tilde s1}, \ldots, \texttt{in}_{\tilde s n}, \texttt{out}_{\tilde sn},). \erasePk{\DD_1}$
\\[2mm]
$\erasePk{
\MainInfer[Sync]
     {\Gamma \vdash \SYNC[\tilde s,n]\{l: P_l\}_{l \in L'} \rhd \Delta, \tilde s: \{l: T_l\}_{l \in L;M,n}@(\pp,n)}
     {\forall l\in L': \DD_l::\Gamma \vdash P_l \rhd \Delta, \tilde s: T_l@(\pp,n)
\quad L' \subseteq L \cup M
\quad M \subseteq L'
     }
  } =
\texttt{out}_{\tilde s \pp} \triangleleft \text{cases}_{L'}; \texttt{in}_{\tilde s \pp} \triangleright \{l: \erasePk{\DD_l}\}_{l \in L'}
$
\\[2mm]
$\erasePk{
\MainInfer[Label]
     {\Gamma \vdash s_k \triangleleft h; P \rhd \Delta, \tilde s: k \oplus \{l: T_l\}_{l \in L}@(\pp,n)}
     {\DD_1::\Gamma \vdash P \rhd \Delta, \tilde s: T_h@(\pp,n) \quad h \in L}
  }  = s_k \triangleleft h; \texttt{out}_{\tilde s\pp} \triangleleft h;\erasePk{\DD_1}$
\\
\normalsize
\center
The other cases are defined monomorphic
\caption{Erasure of Synchronisation from Typing-Derivation}
\label{fig:erasure:process}
\end{figure} 
\begin{figure} 
\small
\vspace{-3mm}
\begin{center}
\begin{eqnarray*}
  \eraseCk{G}_{\tilde s,n,a} &=& a[\pp[n+1]](\tilde s, \texttt{in}_{\tilde s,1}, \texttt{out}_{\tilde s,1}, \ldots, \texttt{in}_{\tilde s, n}, \texttt{out}_{\tilde s, n}).\eraseCk{G}^\star_{\tilde s,n} \\
  \eraseCk{\{l: G_l\}_{l \in L;M}}^\star_{\tilde s,n} 
    &=& \texttt{out}_{\tilde s \pp[1]} \triangleright \{\text{cases}_{L_1 \cup M}: \ldots : \texttt{out}_{\tilde s \pp[n]} \triangleright \{\text{cases}_{L_n \cup M}: \\
    & & \RAND\{\texttt{in}_{\tilde s\pp[1]} \triangleleft l;\ldots;\texttt{in}_{\tilde s \pp[n]} \triangleleft l; \eraseCk{G_l}^\star_{n,\tilde s} \}_{l \in \bigcap_{i=1}^n L_i \cup M}
     \}_{L_n \subseteq L} \ldots \}_{L_1 \subseteq L}
\end{eqnarray*}
\end{center}
\vspace{-3.5mm}
\caption{Conductor Process Generation from a Global Type}
\label{fig:erasure:conductor}
\end{figure} 
\subsection{Erasure Definitions} 
Based on this idea, we translate the synchronisation and symmetric sum types
into the original system \cite{CHY08}, step by step as follows.

\paragraph{Step 1: Process Erasure} 
Only well-typed processes are eligible for erasure, because conductor
processes are generated from the global types. Therefore the erasure
$\erasePk{\cdot}$ is defined on the type derivation in Fig.~\ref{fig:erasure:process} and the
result is the erased process.
We use the notation $\DD::\Gamma \vdash P \rhd \Delta$ to denote a derivation $\DD$
with the conclusion $\Gamma \vdash P \rhd \Delta$.

The case for session request increments the number of participants by one, to make room for the conductor process,
and adds two session channels per user ($\texttt{in}_{\tilde s,\pp}$ and
$\texttt{out}_{\tilde s,\pp}$), for communicating with the conductor.
The conductor process $\eraseCk{G}_{\tilde s,n,a}$ (defined in Step 2) is inserted in
parallel with the resulting session requesting process to ensure it is
available.

The case for synchronisation sends the accepted labels to the conductor, waits
to receive one of the accepted labels and proceeds with the selected branch.


\paragraph{Step 2: Conductor Generation} 
The conductor process $\eraseCk{G}_{\tilde s,n,a}$ was inserted in
parallel with the session requests by the process erasure in Step 1.
The main cases of the conductor generation $\eraseCk{\cdot}$ are in
Fig.~\ref{fig:erasure:conductor}.
Notice that $\eraseCk{G}_{\tilde s,n,a}$ is only a wrapper for
$\eraseCk{G}^\star_{\tilde s,n}$ which prefixes the session
acceptance on channel $a$.  In $\eraseCk{G}_{\tilde s,n,a}$,
$\tilde s$ is the original session channels, $n$ is the
number of original participants, $G$ is the original session type, and $a$ is
the channel the session is created over.

The conductor process generated from a synchronisation receives the accepted
labels from each participant, selects a common label using $\RAND$ and sends
the selected label back to each participant before conducting the chosen
branch.

\paragraph{Step 3: Type Translations} 
\begin{figure} 
\footnotesize
\vspace*{-2mm}
\begin{center}
 \[\begin{array}{rcl}
  \eraseGk{G} &=& \eraseGk{G}^\star_{\keyword{max}(\keyword{pid}(G)), \keyword{max}(\keyword{sid}(G))} \\
  \eraseGk{\{l: G_l\}_{l \in L;M}}^\star_{n,m}
  &=& \pp[1] \to \pp[n+1]: (m+2) \{cases_{L_1 \cup M}: \\
  & & \pp[2] \to \pp[n+1]: (m+4) \{cases_{L_2 \cup M}: \ldots \\
  & & \pp[n] \to \pp[n+1]: (m+2\cdot n) \{cases_{L_n \cup M}: \\
  & & \pp[n+1] \to \pp[1]: (m+1) \{l: \pp[n+1] \to \pp[2]: (m+3) \{l: \ldots \\
  & & 
\pp[n+1] \to \pp[n]: (m+2\cdot n-1) \{l: \eraseGk{G_l}^\star_{n,m} \} \ldots \} 
\}_{l \in \bigcap_{i=0}^n L_i \cup M} \}_{L_n \subseteq L} \ldots \}_{L_1 \subseteq L}
\end{array}\]
\end{center}
\vspace*{-3mm}
\caption{Erasure Mapping for Global Types}
\label{fig:erasure:global}
\end{figure} 
To prove that typability is preserved by the erasure, we define translations of
global types, local types, message types, global type environments and local
type environments to find the types for the result of the erasure.  The main cases for global types are defined in Fig.~\ref{fig:erasure:global}.
The translation $\eraseGk{G}$ of global types is just a wrapper for
$\eraseGk{G}^\star_{n,m}$ where $n$ is the number of participants,
and $m$ is the number of session channels in the original type.

As previously suggested, the symmetric sum is translated to nested branching,
where each participant sends the accepted labels to the conductor, receives
the selected label and continues with the selected branch.

\subsection{Correctness} 
We now prove the correctness of the erasure mapping. 
We start by proving that the typing is preserved, and
the types of the result process is given by the defined type translations.
\begin{THM}[Type Preservation] 
\label{thm:erasure:typable}
If $ \DD::\Gamma \vdash P \rhd \Delta$
then $\eraseGammak{\Gamma} \vdash \erasePk{\DD} \rhd \eraseDeltak{\Delta}$
\end{THM} 
{\sc Proof:} 
By induction on the type derivation $\DD$.
The proof uses a lemma stating that the generated conductor processes are
well-typed.
\\[1mm] 
Next we prove that process congruence ($P \equiv Q)$ is preserved by the erasure.
\begin{THM}[Congruence Preservation] 
\label{thm:fullabstraction:soundness}
\label{thm:fullabstraction:completeness}
\label{thm:ext:5221}
\BREAK
If $\DD_1::\Gamma \vdash P \rhd_{\tilde t} \Delta$ then
for all $Q$ we have that $P \equiv Q$ if and only if there is a derivation
$\DD_2::\Gamma \vdash Q \rhd_{\tilde t} \Delta$
such that $\erasePk{\DD_1} \equiv \erasePk{\DD_2}$.
\end{THM} 
\newcommand{\cin}[1]{\texttt{in}_{#1}}
\newcommand{\cout}[1]{\texttt{out}_{#1}}
\newcommand{\cina}{\texttt{i}_{\pp[1]}}
\newcommand{\couta}{\texttt{o}_{\pp[1]}}
\newcommand{\cinb}{\texttt{i}_{\pp[2]}}
\newcommand{\coutb}{\texttt{o}_{\pp[2]}}
\newcommand{\PC}{\texttt{PC}}
Congruence preservation suggests the erasure preserves semantic properties.
We start by stating the soundness theorem. To do this we define conductors for
partially completed sessions: $\PC(\Delta)$ as the set of possible partial conductor
processes generated from $\Delta$.
By using the partial conductors from the session environment it is now possible
to state the soundness theorem.
\begin{THM}[Soundness] 
\label{thm:erasure:semantics}
$ $
If $\DD::\Gamma \vdash P \rhd_{\tilde t} \Delta$, $P \to P'$,
$\Delta$ coherent and 
$P_C \in \PC(\Delta \circ \Delta'')$ for some $\Delta''$ then there is
a derivation $\DD'::\Gamma \vdash P' \rhd_{\tilde t} \Delta'$ and 
$P_C' \in \PC(\Delta' \circ \Delta'')$ \\
such that $\Delta \to^{0/1} \Delta'$ and $\erasePk{\DD} | P_C \to^\star \erasePk{\DD'} | P_C'$.
\end{THM} 
{\sc Proof:} 
By induction on the derivation of $P \to P'$.
\\[1mm] 
We can extend the above theorem to multiple
steps by induction on the number of steps.
Also the found evaluation of $\erasePk{\DD} \to^\star \erasePk{\DD'}$ performs
exactly the same communication on all non-conductor channels as the original
evaluation $P \to^\star P'$.
\newcommand{\tores}{\nobreak{\ensuremath{\rightharpoondown}} }
\newcommand{\toplus}{\nobreak{\ensuremath{\rightharpoonup}} }

We will now define \emph{conduction steps}, since they play an important role
in formulating the completeness theorem.  This is because all steps performed
by the result of the erasure can be mimicked by the original process up to
\emph{conduction steps}.
A step from $P_1$ to $P_2$ is a conduction step, written $P_1 \tores P_2$ if the step performs
\emph{label selection} or \emph{label branching} on a conductor channel or
unfolding of a \emph{conductor process}; otherwise we write $P_1 \toplus P_2$.
We observe all the extra steps introduced by the erasure are of the form
$\tores$, while the other steps are of the form $\toplus$. Therefore there is a
one-to-one correspondence between the $\toplus$ steps of the erased process,
and the steps in the original process.

\begin{THM}[Semantic Completeness] 
\label{thm:erasure:completeness}
If $\erasePk{\DD_1::\Gamma \vdash P_1 \rhd \emptyset}
\to^\star Q'$ 
then there exists a derivation 
$\DD_2::\Gamma \vdash P_2 \rhd \emptyset$ and $Q$ such that $P_1 \to^\star P_2$ and $\erasePk{\DD_2} \tores^\star Q$ and $Q' \tores^\star Q$.
\end{THM} 
{\sc Proof:} 
By induction on the number of non-conduction steps in $\erasePk{\DD_1} \to^\star Q'$,
using confluence and single-step completeness results.

\paragraph{Healthcare Cooperation (3): Synchronisation Erasure} 
\begin{figure}[t] 
\scriptsize
\vspace*{-4mm}
\begin{tabular}{p{9cm}p{7cm}}
\begin{lstlisting}
P_C' = // Conductor
a[4](d,s,r,in_p,out_p,
     in_d,out_d,in_n, out_n).
out_p :>
{cases_DN: out_d :>
 {cases_DN: out_n :>
  {cases_DN:
    rand {
     in_p <: CaseD;
     in_d <: CaseD;
     in_n <: CaseD;
     out_p :>
     {cases_DN: out_d :>
      {cases_DN: out_n :>
       {cases_DN:
         rand
         {in_p <: CaseDD;
          in_d <: CaseDD;
          in_n <: CaseDD;
          end,
          in_p <: CaseDN;
          in_d <: CaseDN;
          in_n <: CaseDN;
          end },
        cases_D: ... },
       cases_D: ... },
      cases_D: ... } },
   cases_N: ... },
  cases_N: ... },
 cases_N: ... }
\end{lstlisting}
&
\begin{lstlisting}
P_P' = // Patient
a[2..4](d,s,r, in_p, out_p,
  in_d, out_d, in_n, out_n).
out_p<:cases_DN;in_p:>
{CaseD: d<<<eData>;out_p<:cases_DN;in_p:>
 {CaseDD: s>>(schedule);r>>(result);0,
  CaseDN: s>>(schedule);r>>(result);0},
 CaseN: d<<<eData>;out_p<:cases_DN;in_p:>
 {CaseND: s>>(schedule);r>>(result);0,
  CaseNN: s>>(schedule);r>>(result);0} }
P_D' = // Doctor
a[2..4](d,s,r, in_p, out_p,
  in_d, out_d, in_n, out_n).
out_d<:cases_DN;in_d:>
{CaseD: d>><data>;out_d<:cases_DN;in_d:>
 {CaseDD: s<<(eSchedule);r<<(eResult);0,
  CaseDN: r<<(eResult);0},
 CaseN: out_d<:cases_DN;in_d:>
 {CaseND: s<<(eSchedule);r>>(eResult);0,
  CaseNN: r>>(eResult);0} }
P_N' = // Nurse
a[2..4](d,s,r, in_p, out_p,
  in_d, out_d, in_n, out_n).
out_n<:cases_DN;in_n:>
{CaseD: out_n<:cases_DN;in_n:>
 {CaseDD: 0,
  CaseDN: s<<(eSchedule);0},
 CaseN: d>><data>;out_n<:cases_DN;in_n:>
 {CaseND: 0,
  CaseNN: s<<(eSchedule);0} }
\end{lstlisting}
\end{tabular}
\vspace*{-5mm}
\caption{Example Processes after Erasure}
\label{fig:example_erasure}
\end{figure} 
The result of the erasure on the healthcare example from
Section~\ref{sec:typesystem} is shown in Fig.~\ref{fig:example_erasure}.
Since we have shown that the processes from the synchronisation example in
Fig.~\ref{fig:example_sync} are well-typed in
Proposition~\ref{prop:example_sync:typed}, we can apply
Theorem~\ref{thm:erasure:typable} to 
provide $\nobreak{a:\langle \eraseGk{G} \rangle \vdash P'_C \mid P'_P
\mid P'_D \mid P'_N \rhd \emptyset}$. 

As this example illustrates, the result of the erasure does not
capture the nature of the situation in the same way, because it introduces
a conductor process, which is not a natural part of the situation.  It
is not compact either, as the conductor process has $64$ cases. Further we
lose an accurate type abstraction of the dynamics of symmetric
synchronisation, because it is not clear from the encoded type structure
whether it is just a sequence of asymmetric branching actions or
the (intended) atomic multiparty synchronisation, since some of the
key operational structures of the encoding (e.g.~random selection) is
lost in the encoded type.

\subsection{Encodability Criterias}
\label{sec:criterias}
The common properties of encodability from the known separation theorems 
(e.g.~\cite{PalamidessiC:comepsapc}) has been studied \cite{GORLA08}, revealing
a number of desirable criteria. 
Our encoding is {\em type-based}, so we cannot apply this untyped framework
directly. However if we simply change the formulation to use the {\em
type-derivation} instead of the process syntax, our encoding \emph{does}
fulfil the criteria.  

Before we can define and prove the criteria, we need to define the relations
($\asymp_1$ and $\asymp_2$) and properties (successful state) used to define
the criteria.
We select $\asymp_1$ as the process equivalence ($\equiv$), and
define $Q_1 \asymp_2 Q_2$ if and only if $\exists Q. Q_1 \tores^\star Q\ \land\
Q_2 \tores^\star Q$.

\begin{LEM}\label{asymp:congruence}
$\asymp_2$ is a weak barbed
reduction congruence.  
\end{LEM}
{\sc Proof:} 
Immediately $\asymp_2$ is symmetric and reflective by definition. By the
confluence, we can also prove its transitivity.

To define a successful state, we introduce a new process constructor $\surd$,
and extend the typing system to accept $\surd$, and extend the erasure to preserve $\surd$.
A process $P$ is accepting if $P\equiv \surd | P'$ for some $P'$.

We list the new formulation for all the criteria and state the theorem.
For the motivation of each criterion, see 
\cite{GORLA08}. Below, for the sake of readability, we omit $\Gamma$ and
$\Delta$ from the encoding. 

\paragraph{Compositionality criterion} 
{\em For every $k$-ary typing rule ${\textsc r}$ in the typing system of
$\mathcal{L}_1$ and every subset of names $N$ there exists a $k$-ary context
$C_{\textsc r}^N(\__1,\ldots,\__k)$ such that, for all
$\DD_1, \ldots, \DD_k$ with $\textsc{Fn}(\DD_1, \ldots, \DD_k)=N$, it holds
that $\llbracket {\textsc r}(\DD_1,\ldots,\DD_k) \rrbracket =
C_{\textsc r}^N(\llbracket \DD_1 \rrbracket,\ldots,\llbracket \DD_k \rrbracket)$.
}
Note that the information given by derivation (typing) 
in $\DD_1::P_1$ and $\DD_2::P_2$ are essential.  

\paragraph{Name Invariance criterion} 
{\em For every typing derivation $\DD::P$ ($P$ has derivation $\DD$) and name
substitution $\sigma$, it holds that if $\sigma$ is injective, then 
$\llbracket \DD \sigma \rrbracket = \llbracket \DD \rrbracket \sigma'$;
for every $a \in \mathcal{N}$, 
otherwise
$\llbracket \DD \sigma \rrbracket \asymp_2 \llbracket \DD \rrbracket \sigma'$
where $\sigma'$ is such that
$\varphi_{\llbracket \rrbracket}(\sigma(a)) = \sigma'(\varphi_{\llbracket \rrbracket}(a))$.
Here $\varphi_{\llbracket \rrbracket}$ is called the renaming policy and
captures how $\llbracket \cdot \rrbracket$ translates channel names.
} 

\paragraph{Operational Correspondence criterion} 
Let $\to_i$ denote the reduction relation of the system $i$. \\
{\bf (1) Completeness:}
{\em If $\DD_1::P_1$ and $P_1 \to_1^\star P_2$
then there exists a $\nobreak{\DD_2::P_2}$
such that $\llbracket \DD_1 \rrbracket \to^\star_{2}\asymp_2 \llbracket \DD_2 \rrbracket$.}
\\
{\bf (2) Soundness:}   
{\em If $\left\llbracket \DD_1::P_1 \right\rrbracket \to_2^\star Q_1$
then there exists a $\DD_2::P_2$
such that $P_1 \to_1^\star P_2$ and $Q_1 \to_{2}^\star \asymp_2 \llbracket \DD_2 \rrbracket$. }

\paragraph{Divergence Reflection criterion} 
{\em If $\left\llbracket \DD::P \right\rrbracket \to^\omega$ then $P
\to^\omega$} where $\to^\omega$ means infinite reductions.

\paragraph{Success Sensitiveness criterion} 
{\em 
If $\DD::P$ then $P \Downarrow$ if and only if $\llbracket  \DD
\rrbracket \Downarrow$} where  $P \Downarrow$ means $P$ can reach a
successful state.

\noindent Using the above
definition, we arrive at the following main theorem.
\begin{THM} 
\label{thm:criterias}
The erasure mapping satisfies all the encodability criteria. 
\end{THM} 

\Section{Verifying CPG Descriptions}
\label{sec:verification}
\begin{figure}[t] 
\caption{Steps in verifying a CPG description}
\label{fig:verification}
\label{fig:example:processmatrix}
\begin{center}
\quad
\begin{tabular}{p{9.3cm}p{5.7cm}}
\vspace*{-6mm}
\center
{\footnotesize
\begin{tabular}[c]{|c|c|c|c|c|c|}
\cline{3-5}
\multicolumn{2}{c|}{} & \multicolumn{3}{c|}{Roles} & \multicolumn{1}{c}{} \\
\hline
Id      & Name     & Patient & Doctor & Nurse & Predecessors  \\
\hline
1       & Data     & W       & R      & R     &               \\
\hline
2       & Schedule & R       & W      & W     & 1             \\
\hline
3       & Result   & R       & W      & N     & 2             \\
\hline
\end{tabular}
}
\vspace*{-6mm}
&
{\bf Process Matrix} \cite{pm08example} \newline
Formal representation of CPGs
\vspace*{-5mm}
\\
\center\includegraphics[width=4.5cm]{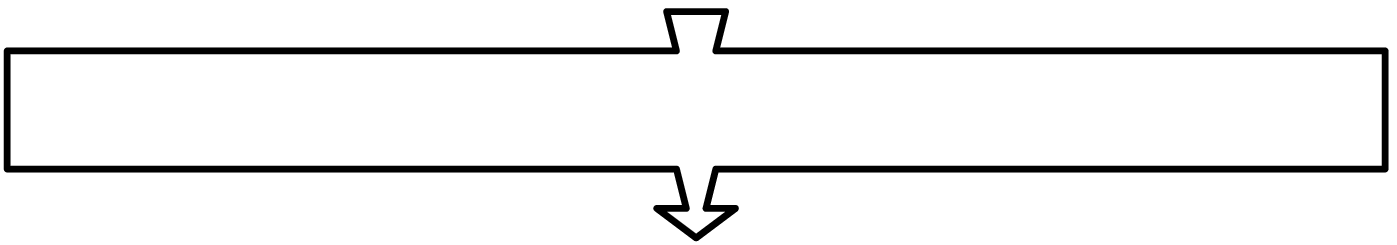} & \\
\vspace*{-13mm}
\center{\bf Process Matrix Encoding} &
\\
\vspace{-10mm}
\scriptsize
\begin{lstlisting}
{^Pdata: 1=>2:2<String>;1=>3:3<String>;rec stateD. 
  {^Pdata: 1=>2:2<String>;1=>3:3<String>;stateD,
   ^Dschedule: 2=>1:1<String>;2=>3:3<String>;rec stateDS.{...}
   ^Nschedule: 3=>1:1<String>;3=>2:2<String>;rec stateDS.{...}
} }
\end{lstlisting}
\vspace*{-6mm}
&
\vspace*{-5mm}
{\bf Global Type}
\\
\vspace*{-7mm}\center\includegraphics[width=4.5cm]{gfx/arrow_translation} & \\
\vspace*{-13mm}
\center{\bf Type Projections} &
\\
\vspace{-10mm}
\scriptsize
\begin{lstlisting}
{^Pdata: 2!<String>;2!<String>; rec stateD. 
  {^Pdata: 1=>2:2<String>;1=>3:3<String>;stateD,
   ^Dschedule: 1?<String>;rec stateDS.{...}
   ^Nschedule: 1?<String>;rec stateDS.{...}
} }
\end{lstlisting}
\vspace*{-5mm}
&
\vspace*{-5mm}
{\bf Local Types}
\\
\vspace*{-8mm}\center\includegraphics[width=4.5cm]{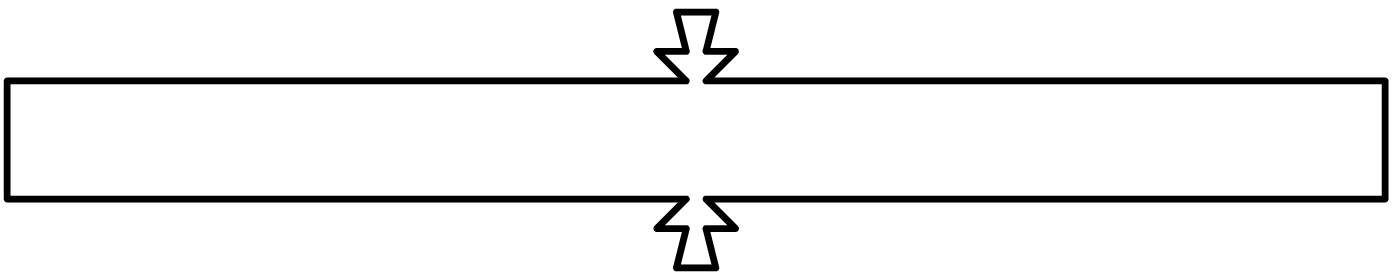} & \\
\vspace*{-13mm}
\center{\bf Verification}
&
\vspace*{-10mm}
\\
\vspace{-14mm}\center
\scriptsize
\begin{lstlisting}
sync((p,d,n),3)
{^Pdata: s[2]<<<e>;s[3]<<<e>;def StateD(s)=sync((p,d,n),3)
  {^Pdata:s[2]<<<e>;s[3]<<<e>;StateD(s),
   ^Dschedule:s[1]>>(x);def StateDS(s)=...,
   ^Nschedule:s[1]>>(x);def StateDS(s)=...
  } in StateD(s) }
\end{lstlisting}
&
\vspace*{-5mm}
{\bf Implementations}
\end{tabular}
\end{center}
\vspace*{-5mm}
\end{figure} 
\noindent This section describes how symmetric sum types can verify
implementation conformance to a CPG \cite{cpg09} described using the
Process Matrix. The verification is performed by three steps 
in Fig.~\ref{fig:verification}, as illustrated below.

\paragraph{Process Matrix.} 
The Process Matrix representation consists of a table with one row for
each action.  Each row has a number of columns: The \textbf{Id} and
\textbf{Name} columns are used to identify the action, and the
\textbf{Predecessors} column holds the \emph{Ids} of the actions the action
depends on. Before an action can be executed its predecessors must have been
executed. If all the predecessors of an action have been executed we say that
the action is {\em executable}.  Finally there is one column for each participant
(called roles), where the content is either
\verb|R| meaning the participant can read the action-data but not execute it,
\verb|W| meaning the participant can execute the action and read its data or
\verb|N| meaning the participant cannot execute the action or read its data
(see \cite{pm08example} for a more adequate description).
The Process Matrix in Fig.~\ref{fig:example:processmatrix}
describes the scenario from the introduction, 
except that the patient automatically gives the
data to both the doctor and the nurse, and the user 
can perform the actions multiple times (by an implicit recursion), 
until all the actions are executed. 

\paragraph{Process Matrix Encoding} 
\label{sec:verification:matrixencoding}
Any CPG in a Process Matrix can be encoded as a global type
automatically.  We explain this encoding by translating the above
Process Matrix example.  In the resulting type, the
state is described by the set of actions that have been executed,
leading to a finite but exponential number of states.  The
representation of each state (except the completed state) is a
symmetric sum with one branch for each role that can execute each
executable action. The content of each branch consists of the
executing participant sending the created data to all other
participants with read or write access, followed by the state where
the executed action is added, and depending actions have been removed.

Parts of the global type is included in Fig.~\ref{fig:verification}.
Notice that the resulting type uses recursion: this is to describe
an implicit recursion in the Process Matrix where the state reached
after an action does not have to be a new state, but can be the same
as the state before the execution of the action, or even from previous steps.
This is the case for the above example if the data is sent, the
appointment is scheduled, and then the data is resent. The resulting
state would then be the state where only the data action has been executed,
which is the same as the second state.  The described method can be
extended to translate any Process Matrix into a global type.

The conversion of CPGs from the Process Matrix, to session type allows the data
to be exchanged directly between the participants, while the current
implementations rely on a centralised database for the exchange.
This means the translation offers a distributed implementation of the Process
Matrix, which has not been known before.
A formally defined symmetric global synchronisation primitive, together with
its type discipline and encodability, offers a firm basis for such
implementations.

\paragraph{Projection and Verification} 
When we have created the global type expressing the CPG, a process implementing
one of the participants can be verified to conform with the workflow, by
projecting the global type to the local type of that participant, and typechecking
the process against the local type. Parts of the local type and the process for
the Patient are described in Fig.~\ref{fig:verification}.

\paragraph{Generalisation} 
We have now described how to use the multiparty session types extended with
symmetric sum, to express CPGs formalised using the Process Matrix.
We believe many other workflow frameworks (such as large parts of the BPMN)
can be encoded as multiparty session types with symmetric sum, and this would
allow the type-system to serve as a common representation, enabling interaction
between different frameworks and implementing features (such as automatic
user-interface generation) only for symmetric sum types, and apply it to all
the encoded frameworks.

\subsection{Implementation}
\label{sec:implementation}
We have created an \verb|ascii| syntax for the asynchronous
\picalc with multiparty sessions and symmetric synchronisation called
\textsc{apims}, and implemented a typechecker and an interpreter. This is to
our knowledge the first prototype implementation of the $\picalc$ with
multiparty sessions and multiparty session types.  The implementation along
with example programs can be found on the \textsc{apims} website
\cite{apims:projectpage}.

The implementation extends the calculus with a \keyword{guisync}
constructor to support user interaction via GUIs. The
\keyword{guisync} is the result of extending the
\keyword{sync} for user input. Each label has a set of
typed arguments that must be given using the GUI before that choice is
accepted, and the given arguments can be used by the process in that
branch. This simple extension allows the processes to implement GUIs
and the type system guarantees that
the GUI for each participant will respect the protocol, hence
the workflow. The mandatory labels ensure that the GUI must allow all
the users (the people using the interface for each participant) to agree in
each synchronisation, thus avoiding the GUIs causing a disagreement w.r.t. the
theory of a symmetric synchronisation.

The GUI shows the received data, the choices offered by the process, input
fields for the data needed for each choice, and buttons to accept/reject each
choice.  Fig.~\ref{fig:example:states} shows three screen-shots, displaying the
doctor's GUI for each state and how each choice affects the state.  As soon as
all the participants of a session accepts the same choice, the processes
continue with the accepted branch.  The GUI implementation for each participant
can be created automatically from the Process Matrix.

The original implementation of the Process Matrix called \emph{Online
Consultant} by \emph{Resultmaker} \cite{pm08example} is database based.  This
means that communication consists of the sender uploading information to the
server, and all participants must query the server when using the information.
Implementing the workflows using the $\picalc$ and session types not only gives
the \emph{Process Matrix} a formal semantics, but also allows an implementation
where participants communicate their data as peer-to-peer. This offers more
natural and robust realisation of the workflows, and relieves the system from
the server bottleneck.
\begin{figure} 
\hspace*{1mm}
\centering
\includegraphics[width=15cm]{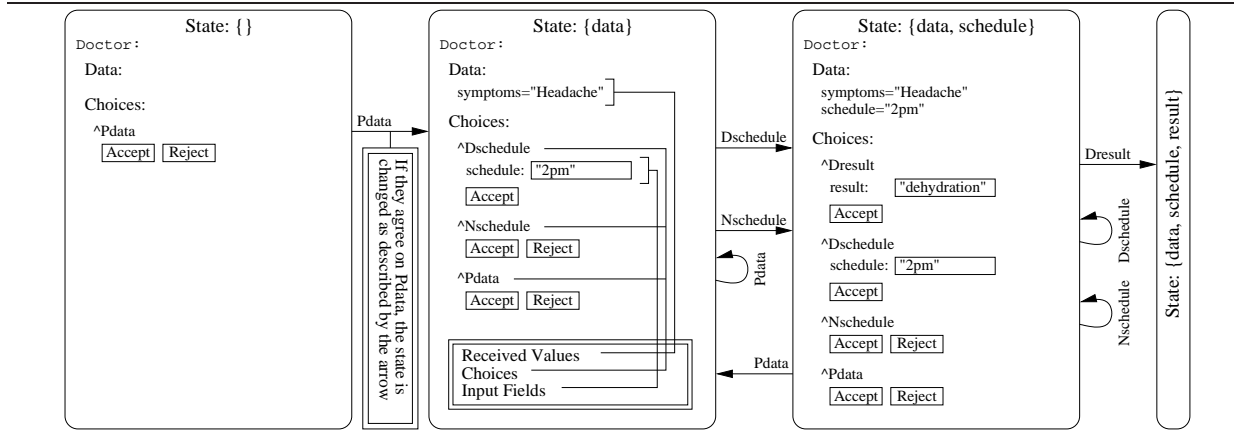}\label{fig:gui:structure} \\
\caption{States and screenshots for the doctor GUI}
\label{fig:example:states}
\end{figure} 

\Section{Related and Future Work}
\label{sec:relatedwork}
There are existing studies on self/broadcast synchronisations
\cite{HoareC:ComSP,prasad01}.  The symmetric sum proposed in the
present paper is different because it allows all the participants to
influence the choice equally and, to formulate this notion adequately,
demands a session-based operational framework. Another 
difference is the use of the type discipline to control this
complex synchronisation framework, which is not found in
the foregoing work. Note that the type discipline allows multiparty
progress and communication-safety for participants, which is not
generally ensured in existing untyped self/broadcast synchronisation
primitives. Our primitive and its type-checker 
are applicable not only to 
Process Matrix, but also multiparty synchronisations in general with 
strong safety guarantees. 

The symmetric synchronisation is similar to the consensus in Weak Byzantine Agreement (WBA)
\cite{F83,L83,ADG84,FL85} which is a formalisation of the database commit problem.
The similarity is that a number of processes need to end up with a common choice.
In contrast to symmetric sum, WBA only has two possible choices (0 and 1). 
Not all participant has to initially accept the final decision, 
but if all processes agree 
initially, the result should be the initial preference.
WBA is studied in an untyped settings on unreliable networks, with faulty
processes (with arbitrary behaviour). 

The symmetric sum is also similar to the symmetric choice $\square$ in CSP and
the mixed choice in the \picalc \cite{PalamidessiC:comepsapc}. The main
difference is these preceding primitives are restricted to two party
synchronisations.  Our result is consistent with the non-encodability of the
{mixed-choice} \picalc in the {separated choice} \picalc
\cite{PalamidessiC:comepsapc}: our erasure is defined on {\em typing
derivations}, and cannot be made homomorphic on \emph{processes}.  For example,
take $P = (\nu a)(P_1 | P_2)$ where 
\begin{quote}
$P_1 = \overline{a}[2](s).\mathtt{sync}\{\underline{l1}: P_{11}, l2: P_{12}
\}$ and $P_2 = a[2](s).\mathtt{sync}\{\underline{l1}: P_{21}, l3:
P_{23}\})$.  
\end{quote}
This process shows that the erasure cannot be interpreted
as an encoding from processes $\llbracket \cdot \rrbracket$ where
$\llbracket P_1 | P_2 \rrbracket = \llbracket P_1 \rrbracket |
\llbracket P_2 \rrbracket$, because the result of $\llbracket P_1
\rrbracket$ depends on the context $P_1$ is in: the conductor inserted
by the second step of the erasure depends on the type of $a$ which
depends on the other process.  In the given context, the conductor
must consider the labels $l1,l2$ and $l3$, and this could not be
generated from $\llbracket P_1 \rrbracket$ because $P_1$ does not
contain any information about $l3$.  As noted above, the symmetric sum
and synchronisation construct differs from the mixed choice and from the
untyped asymmetric, directed sums whose encodability is studied in
\cite{NestmannU:decce,Nestmann00}, in that it is multi-party
synchronisation for a fixed number of participants ensured by
the underlying session type discipline. 


Types for the multiparty interactions are studied in the
\emph{conversation calculus} \cite{conversationtypes09} and
\emph{contracts} \cite{cp09}.  The former has choice behaviours where
the channel-based communication is replaced by conversation
environments allowing multiple participants, while the latter
uses a process-based specification of protocols relying on
internal and external choices, where conformance is formalised based
on must preorder (so that we can ensure liveness).  Our implementation
crucially relies on the choreographic description based on global
types: in particular, global types can offer a tractable, clear
type-directed generation from the Process Matrices as described in
Section~\ref{sec:verification}.
\label{sec:futurework}

As future work, we plan to extend our work 
with logical assertions based on \cite{bhty09} 
in order to describe and ensure the
communicated data fulfil desired properties (for example, ``the prescribed medicine
doses are less than the lethal amount'').
With the assertions, we can add arguments (state) to the recursive
types, and conditions to the branches in a choice, so that it will
lead to a more efficient generation from the Process Matrix. 

\subsection*{Acknowledgements}
The first author is supported by 
the \emph{TrustCare} project, funded by the Danish Strategic Research
Agency, Grant \#2106-07-0019. The last two authors are partially 
supported by EPSRC EP/F003757, EP/F002114, EP/G015635 and EP/G015481. 

\bibliographystyle{abbrv}
\bibliography{main}
\end{document}